\documentstyle[aps,epsf,twocolumn]{revtex}
\begin{document}
\draft
\wideabs{
\title{ Characteristic Spatial Scales in
	Earthquake Data}
\author{G. Z{\"o}ller, R. Engbert, S. Hainzl and J. Kurths}
\address{Institute for Theoretical Physics and\\ Astrophysics 
         at the University of Potsdam\\
	D--14415 Potsdam, Germany\\
	email: gert\symbol{64}ik.uni-potsdam.de}
\date{\today}
\maketitle
\begin{abstract}
We present a new technique in order to quantify 
the dynamics of spatially extended systems. Using a test on the
existence of unstable periodic orbits, we identify intermediate spatial
scales, wherein the dynamics is characterized by maximum nontrivial
determinism. This method is applied to earthquake catalogues containing
time, coordinates and magnitude. As a result we extract a set of
areas with significant deterministic and low--dimensional dynamics from the
data. Finally, a simple model is used to show that these scales can be 
interpreted as local spatial coupling strengths. 
\end{abstract}
}

\section{Introduction}\label{sec:intro}
In recent years a lot of techniques to analyse single complex 
time series have been developed (cf.~\cite{Graetal91}). But a special
challenge is the analysis of spatiotemporal dynamics, especially of
natural systems.
Although such systems become more and more important,
e.g. in environmental research~\cite{Hasetal94}  or brain imaging
techniques~\cite{Hametal93}, 
little is known about how to analyse the 
corresponding data. Mostly the spatial extension is not taken into
account and the more--dimensional data are simplified to
one--dimensional time series~\cite{Ros94}. 
This may be senseful for systems, which behave approximately homogeneously
in space, but in general this assumption is not fulfilled for natural
systems. 
As a
consequence much information is lost after the modification
of the data, e.g. by describing a complex spatial pattern by a
single number 
like a mean value or a variance or a more complicated parameter like a 
fractal dimension. More dynamical approaches are 
based on a 
decomposition of spatiotemporal patterns into special basis functions,
e.g. wavelet transformation~\cite{Chu92} or the Karhunen--Lo\`{e}ve
method~\cite{Frietal92}. But these techniques are not appropriate 
in the case of rather irregular and noisy data.

The purpose of this 
contribution is to search for characteristic 
spatial scales, on which the interesting dynamical properties of the
system can be observed. 
This idea allows to take the spatial extension of a system into
account and is applicable to a large variety of data. 
Analysing the
dynamics on these scales will give 
a maximum of nonlinear and low--dimensional
determinism~\cite{Peietal96}. As a result the underlying dynamics can be represented locally
by a vector in a low--dimensional space.     
This idea was suggested by Rand and Wilson~\cite{Ranetal95} for
systems, where the global dynamics is in a steady state and, 
therefore, trivial in the 
thermodynamic limit. 
Most natural systems, however, are far from thermodynamic
equilibrium~\cite{Hak88} and their system size is far from infinite size. As
a consequence, we observe complex dynamical behavior even on the
largest observable scales. Therefore, the concept of Rand and Wilson
has to be modified for the analysis of this class of systems. 

On small scales the dynamics is
dominated by intrinsic 
stochasticity and on large scales  spatial averaging over dynamically
desynchronized parts of the system suppresses determinism as well as
stochastic fluctuations. Therefore, the averaging procedure itself
may be exploited for the analysis of spatiotemporal time series.  
We search for intermediate scales, where,   
on the one hand a deterministic signal is observed and  on the other
hand the loss of dynamical information 
arising from spatial averaging is as small as possible. 
To identify such a  nontrivial determinism, we use a formalism of Pei
and Moss~\cite{Peietal96} based
on the existence of unstable periodic orbits. 

As a dynamically rich and interesting subject, we apply this approach
to natural 
seismicity which provides a lot of 
chaotic and fractal features, because the underlying processes like
stress accumulation and ruptures are strongly nonlinear (cf.
~\cite{Mei94,Tur92}). Here we mention  
the famous Gutenberg--Richter law~\cite{Gutetal54}
\begin{equation} log N = am+b, \label{eq:gr}\end{equation} where $N$ is the number of
earthquakes with  
magnitude greater than or equal to $m$. The
magnitude $m$ is related to the seismic energy $E$ by
\begin{equation} log E = cm+d \label{eq:energy}\end{equation} with constants $a$, $b$, $c$ and $d$.
Inserting Eq.~(\ref{eq:energy}) into Eq.~(\ref{eq:gr}) a power--law for the
density function $D(E)$ is obtained
\begin{equation} D(E)=\frac{dN}{dE}=\frac{a}{c}\ 10^{\ b-\frac{ad}{c}}\ 
E^{\ \frac{a}{c}-1}=C\ E^{\tau}.\label{eq:power}\end{equation} In
general power--laws 
with noninteger exponent $\tau$ indicate fractal
distributions~\cite{Tur92}.    

Due to technical difficulties, the analysis of earthquake data is not
straightforward. These data are point--like and not equidistant in
time and space, of course. Furthermore most catalogues are not complete in the
sense that they contain all micro--quakes. In the next section
we describe a pre--processing of the catalogue in order to generate
regular data.   In Sec.~\ref{sec:scales} we give the algorithm for the
extraction of the scales. This algorithm is applied in Sec.~\ref{sec:datan}
to the real data and in Sec.~\ref{sec:model} to the model
data. Finally, we summarize the main ideas and results of our work.

\section{Data and Pre--processing}\label{sec:data}
The data, which are investigated here 
incorporate 10.779 earthquakes recorded in Armenia between 1974 and
1994~\cite{GFZ94}.   
Each earthquake is described by a four--dimensional
vector consisting of time, longitude, latitude and magnitude. 
The
events are distributed very complex, i.e. not equidistant, in 
time and space. 
For the data analysis  it is helpful to handle with 
data, which are at least equidistant in time. Therefore, we subdivide
the time into 
intervals $T_i=i\cdot \Delta t$ and the space into cells $A_j$ where
the local dynamics is considered; the shape of the spatial cells 
will be determined in the next section. 
Corresponding to Eq.~(\ref{eq:energy}) we introduce 
the total energy $E_{ij}$ in this cell by 
\begin{equation} 
E_{ij} = \sum\limits_{t\in T_i,\vec x\in A_j} 10^{m(\vec x,t)}.
\end{equation}
For simplicity we have defined $c=1$ and $d=0$ for the constants in 
Eq.~(\ref{eq:energy}). 
Now we can define an effective magnitude $M_{ij}$ as a continuous
function of time and space by 
\begin{equation} M_{ij} = log E_{ij}. \label{eq:meff}\end{equation}
The subdivision into cells has to be chosen fine enough so that the loss
of dynamical information is as small as possible. 

\begin{figure} 
\begin{center}
\epsfxsize=8cm
\epsfysize=8cm
\epsfbox{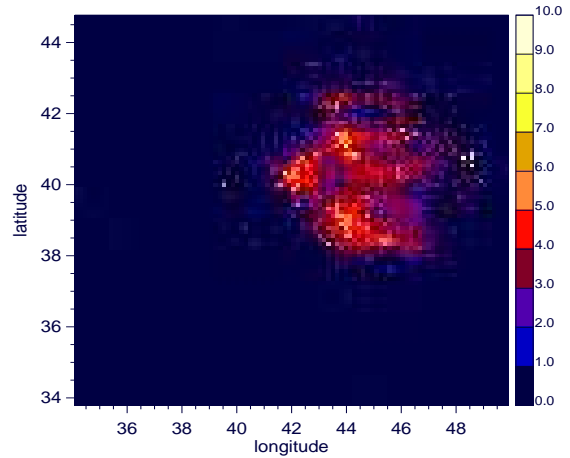}
\end{center}
\caption{Spatial distribution of seismic activity (colour coded).}
\label{sigtot}
\end{figure}

For a fixed spatial area
$A$ around a point $\vec x_f= $ (latitude,longitude) we compute the function
\begin{equation}
\sigma_{A,\vec x_f}(t) =\int\limits_0^t\ \left( M_{A,\vec x_f}(t')
-{\overline M_{A,\vec x_f}}\right)\ dt',
\label{eq:intsigma}
\end{equation}
where ${\overline M_{A,\vec x_f}}$ is a sliding temporal mean of $M_{A,\vec
x_f}(t)$. 
We have found that $500$ time intervals (corresponding to an interval
length of 15 days) and a window length of $100$
for the sliding mean are appropriate values for our calculations. 
In Fig.~\ref{sigtot} we show the
spatial distribution of seismic activity
 $\Sigma_j$ integrated over the whole time $[0,t_{max}]$
\begin{equation} \Sigma_j = \log\int\limits_0^{t_{max}}\ 
\sum\limits_{\vec x\in A_j} 
 10^{m(\vec x,t)}\ dt.\end{equation}

\section{Characteristic Spatial Scales}\label{sec:scales}
The challenge here is to include the spatial extension of
the system in the analysis of the earthquake data. For this aim we
want to extract spatial regions from the data, wherein the dynamics
is highly deterministic. In contrast to
global descriptions,  
where the spatial extension is simplified to a number, 
much less dynamical information is lost then.
 
The main idea is to search for deterministic dynamics in time 
series $\sigma_{A_j,\vec x}(t)$ corresponding to a fixed point $\vec x$ 
and different areas $A_j$.  

We call a scale {\em characteristic}, if the dynamics in one area is
deterministic with higher significance than in the 
other ones.  
The time series $\sigma_{A,\vec x}(t)$ corresponding to this scale 
is then the appropriate choice for further
investigations of the local dynamics. In our calculations we use
circles with increasing radius for these areas. 

We describe now the procedure for the extraction of characteristic
scales in four steps:

\begin{enumerate}
\item[i.] Scatter plot of the time series $\{\sigma_n\}$:\\
For the scatter plot we choose an embedding of $\{\sigma_n\}$: 
\begin{equation} f(\sigma_n) = \sigma_{n+k}. \end{equation} 
The lag $k$ is determined by the condition that the auto--correlation of the
time series is sufficiently small. 
This is fulfilled for $k=4$.  

\item[ii.] Detecting candidates for unstable periodic orbits (UPOs):\\
The identification of unstable periodic
orbits~\cite{Peietal96,Schietal94} rests on
the occurrence of a special sequence of points in the time series. If
the system's trajectory 
enters such a sequence, an UPO is visited by following a predictable
pattern of values of the time series.
An UPO is an intersection point of a stable and an unstable
manifold~\cite{Peietal96}.  
In the vicinity of an UPO, we can approximate these manifolds locally
linear. 
We define the following criteria for an UPO candidate 
(see Fig.~\ref{sequence}): (1) the UPO
itself is close to the line of identity: the perpendicular distance to the
line of identity  is smaller than the mean of the perpendicular 
distances of the five points and (2)
a straight line approaches the line of identity (stable manifold) and
a straight line diverges from the line of identity
(unstable manifold). 
We want to point out that these conditions are
only capable to detect candidates for UPOs in time series, because they
are necessary but not sufficient ones. We refer to~\cite{Soetal96} for a 
method which is more appropriate to detect UPOs itself. 

\begin{figure} 
\begin{center}
\epsfxsize=8cm
\epsfysize=6cm
\epsfbox{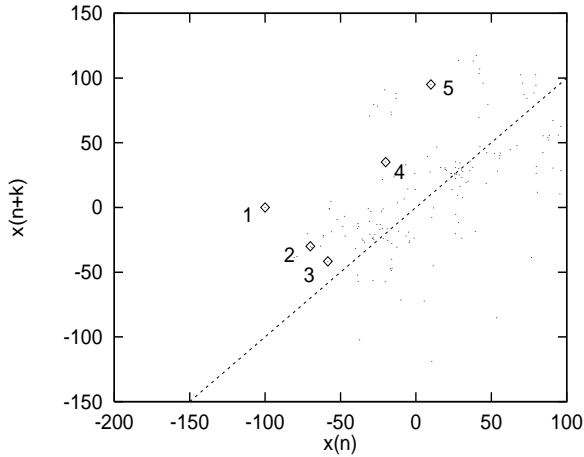}
\end{center}
\caption{Visual definition of an UPO candidate in a time series
$x(n)$: 
the dots are the data points, the
squares denote the characteristic sequence of points (1,2,3: stable
direction; 3,4,5: unstable direction), 
point 3 is the
UPO candidate which is close to the line of identity (dotted line).}
\label{sequence}
\end{figure}

\item[iii.] Definition of the statistical significance:\\
 The statistical significance for the number
of UPOs can be defined by comparing the original data with
a large number of surrogate data files. Therefore, we count the number
of UPOs in the original data, $N, $ and in the surrogate data, $N_s, $ and
compute the statistical 
significance~\cite{Peietal96}: 
\begin{equation} K = \frac{N-\langle N_s\rangle}{\sigma_s}, \label{eq:sig} \end{equation}
where $\sigma_s$ is the standard deviation of $N_s$ and $\langle N_s\rangle$ the
mean value of $N_s$ over all surrogate data files. 
For Gaussian distributions $K\ge 3$ is equivalent to a confidence
level of $99\% $ to reject the null hypothesis that the original data 
are linear in the sense that they do not contain  a significant number
of UPO candidates. 
Our surrogate data are generated by phase randomization
and amplitude adjustment of the original data~\cite{Theetal92}. This
guarantees that the auto--correlation function as well as the distribution of
the data are conserved.

\item[iv.] Extraction of characteristic scales:\\
The procedure to  extract the characteristic scales from the data works 
as follows: One point in space is surrounded by circles
with increasing radius: $r=\Delta r,\ 2\Delta r,\ \dots\ $. The step
width is chosen to $\Delta  r = 5 $ km. 
For each circle a time series $\sigma_n$ is generated
from the data as described in Sec.~\ref{sec:data}.  In these time series 
we detect the
UPO candidates and compute the significance $K(r)$ in comparison with $100$
surrogate data files.
A  characteristic scale  around a point $\vec x$ with 
radius $r_{max}$ and significance $K_{max}$ is given, if
(1) $K(r_{max})\ge 3$ and (2) 
$K(r_{max})=K_{max}=\max\limits_r K(r)$.  
This procedure is applied to each point of a  $20\times 20$ lattice. 
To avoid finite size effects we exclude boundary points in
space. Note also that the number of events in the boundary regions is
too small to provide reliable results. 
\end{enumerate}

We have checked that the results do not depend sensitively on the 
parameters.


\section{Results of data analysis}\label{sec:datan}
\begin{figure} 
\begin{center}
\begin{minipage}{0.48\textwidth}
\begin{center}
\epsfxsize=8cm
\epsfysize=6cm
\epsfbox{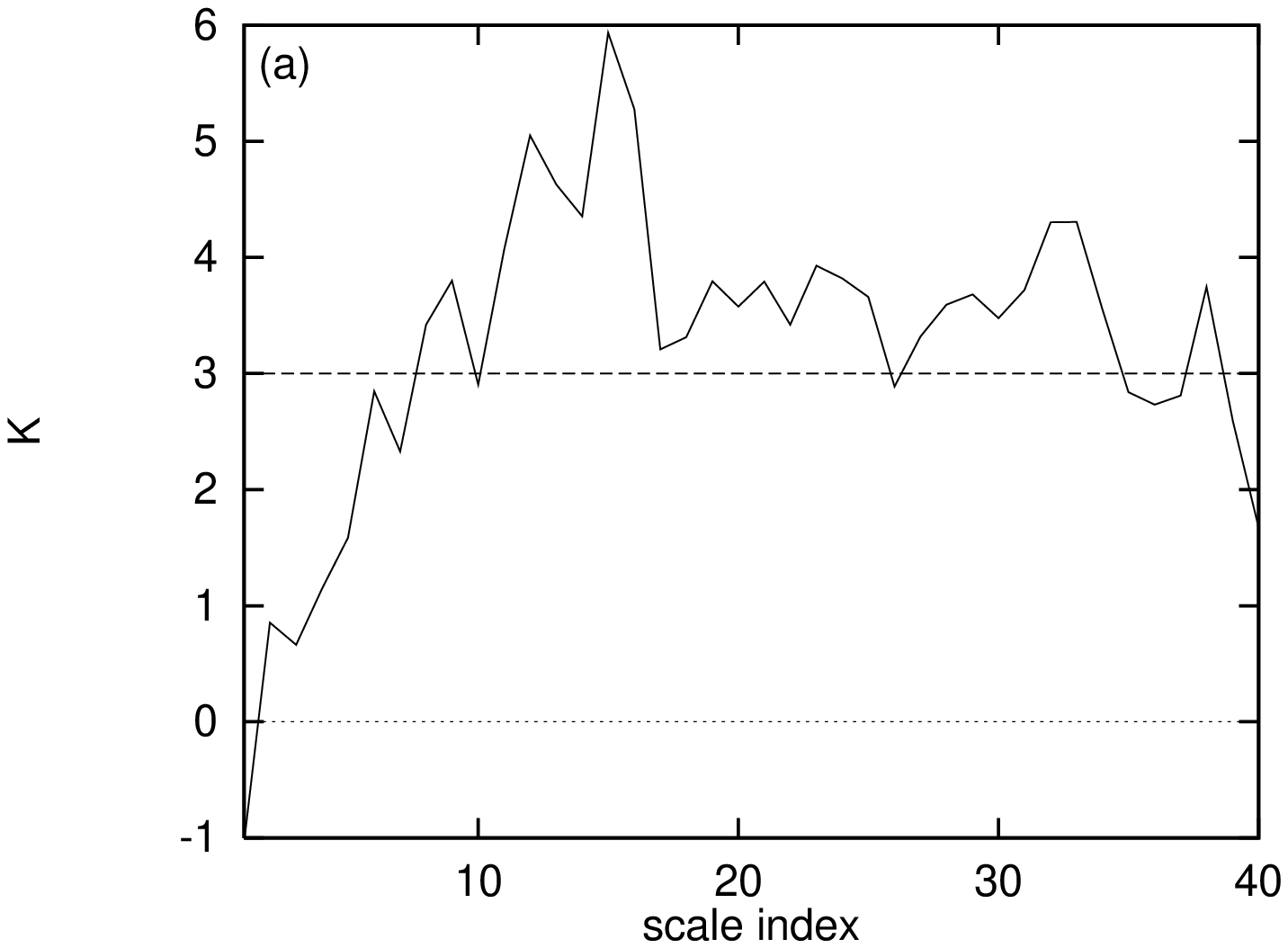}
\end{center}
\end{minipage}
\begin{minipage}{0.48\textwidth}
\begin{center}
\epsfxsize=8cm
\epsfysize=6cm
\epsfbox{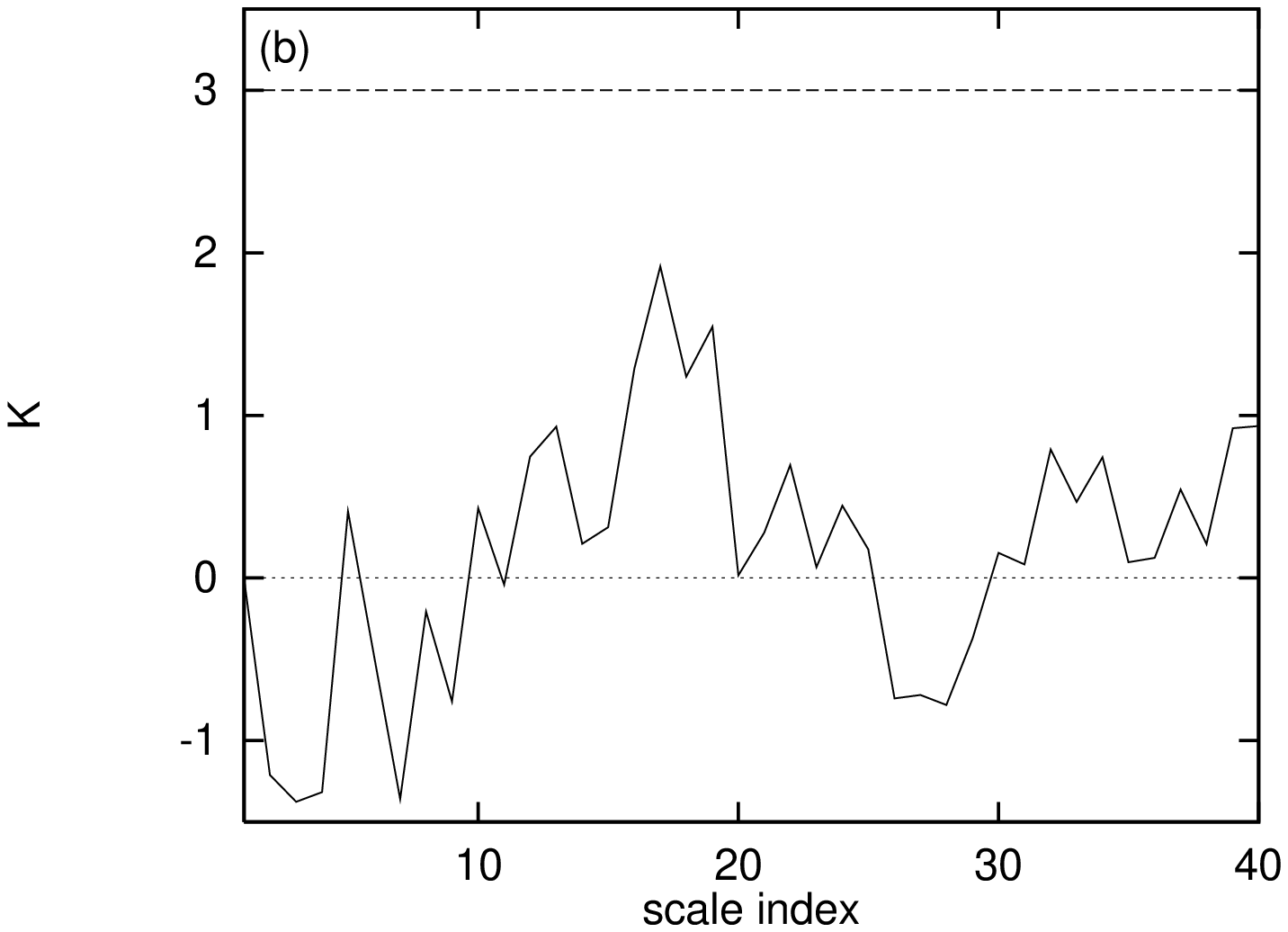}
\end{center}
\end{minipage}
\end{center}
\caption{Significance $K$ as a function of the spatial scale for two
points in space: (a) $\vec x=(40.3,41.8), $ (b) $\vec
x=(41.4,43.7).$ A unit of the scale index is 5 km.}
\label{sig.arm}
\end{figure}

Applying our algorithm to the earthquake data exhibits indeed in some
regions such characteristic scales. Two examples are shown in
Fig.~\ref{sig.arm}, where 
we have plotted the significance $K$ from Eq.~(\ref{eq:sig}) as a
function of the spatial scale represented by an index for two fixed points in 
space. 
In Fig.~\ref{sig.arm}(a) we
observe values for $K$ up to six and a clear maximum for the scale
index 15. Here it is possible
to assign a characteristic scale to the point $\vec x$. 
In contrast to this we
cannot find such a significant scale in Fig.~\ref{sig.arm}(b).  
The level of $K=3$ is not reached here. 
Note that the significances in Fig.~\ref{sig.arm}(a) and
~\ref{sig.arm}(b) are quite different, although the points $\vec x$ are
very close in space. This result underlines that an averaging over
the whole space or large parts of the space is indeed questionable.  

However, the rule for the extraction of the characteristic scale is
rather simple. Not for all points a clear maximum of the significance
can be observed. 
In some cases there are two or more maxima, or the shape of
the curve $K(r)$ is approximately constant. For the 
investigation  of the local dynamics the function $K$ should be studied in
detail. But as a global approach, the rules for the selection of the
scales seem to be reasonable.  

\begin{figure} 
\begin{center}
\epsfxsize=8cm
\epsfysize=8cm
\epsfbox{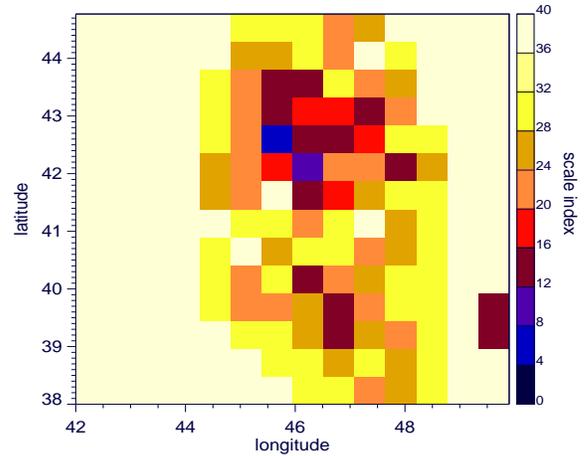}
\end{center}
\caption{Distribution of characteristic scales in space (colour
coded); a unit of the scale index is 5 km.}
\label{scales}
\end{figure}

The distribution of the scales in space is given in
Fig.~\ref{scales}. 
The scale sizes are related to the general seismic activity
(see Fig.~\ref{sigtot}); the dynamics in very active regions is
mostly 
determined on small scales, whereas in regions with less activity 
larger scales dominate. 
It is important to check, if this relationship is linear. In this case
the distribution of the scales would simply reflect the seismic activity
given in Fig.~\ref{sigtot} and the results were completely trivial.
This is, in fact, not the case, because the linear correlation
coefficient between these quantities is almost zero.   

Our technique uses unstable periodic
orbits to quantify nonlinear determinism. Next we check, 
whether a simpler discriminating statistic can be used for this aim. 
Therefore, we perform the same algorithm to obtain the time series
(Eq.~(\ref{eq:intsigma})),  
and compute then the 
skewness~\cite{Theetal92} of these series 
instead of numbers of UPO candidates. The skewness is a rather simple
quantity, which 
indicates  nonlinear behavior and time reversal invariance in time
series. Comparing original data and surrogate data by
Eq.~(\ref{eq:sig}), 
no significant deviation is observed. As a consequence, we need indeed a
more powerful method to detect nonlinear features in the dynamics.

\section{Calculations with model data}\label{sec:model}
To connect our results of data analysis with physical properties of
the system, we want to apply  our technique also to standard models of
seismicity. 
In this way we can control the properties of the simulated data 
by changing some parameters of the model. 

\begin{figure} 
\begin{center}
\epsfxsize=8cm
\epsfysize=6cm
\epsfbox{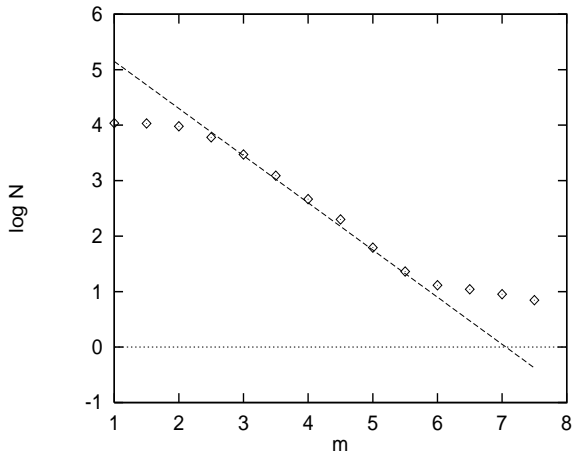}
\end{center}
\caption{The Gutenberg--Richter law (Eq.~(\protect\ref{eq:gr})): the
points denote the number of earthquakes with magnitude greater or
equal to $m$, the dashed line is a linear fit for $2\le m\le6.$ }
\label{arm}
\end{figure}

An important point is the incompleteness of the earthquake
catalogue, which may have influence on the results. For instance the 
Gutenberg--Richter law is only fulfilled for $2\le m\le
6$ (see Fig. ~\ref{arm}). In the region $m>6$ the number of events is
too small to provide a good statistics and for $m<2$ a large 
number of micro--quakes is not measured for this catalogue. 
To analyse the dependence of the significance of UPOs on the
completeness of the data, we generate a synthetic earthquake catalogue 
and assume that these surrogate earthquake data behave 
similar to the real data.

There is a standard model for earthquakes 
by Bak and Tang~\cite{Baketal89} exhibiting self--organized 
criticality. This is a cellular automaton analogue to
well--known stick--slip--models~\cite{Buretal67}, which 
describes the earth's crust to be in a stationary critical state so
that the distribution of earthquakes follows the   
Gutenberg--Richter relation.
For a detailed description of the model we refer to~\cite{Baketal89}.
Here we only recall the rules for the redistribution of energy $Z(i,j)$, if
a cell $(i,j)$ is in a critical state, i.e. $Z(i,j)> Z_{crit}$ :
\begin{eqnarray} 
\label{eq:bak}
\lefteqn {Z(i,j)\longrightarrow Z(i,j)-4}\\
& Z(i\pm 1,j\pm 1)\longrightarrow Z(i\pm 1,j\pm 1)+1 & \nonumber 
\end{eqnarray}

\begin{figure}
\begin{center}
\epsfxsize=8cm
\epsfysize=6cm
\epsfbox{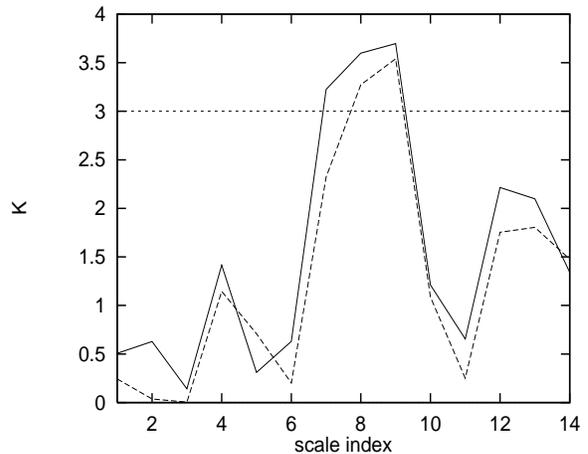}
\end{center}
\caption{Significance $K$ as a function of the spatial scale for the 
point $\vec x=(17,18)$ in the model of Bak and Tang; solid line:
truncated model catalogue, dashed line: complete model catalogue.}
\label{bakscale}
\end{figure}

With a $50\times 50$--grid and the critical value $Z_{crit}=3$ we
create two earthquake catalogues: (A) a ``complete'' catalogue including
all micro--quakes and (B) a ``truncated'' one, where quakes with
$E<10$ ($m<1$) are discarded. We choose the size of (2) equal to that
of the real data. In Fig.~\ref{bakscale} we compare the results of
(A) and (B)  
for one point in space; like in Fig.~\ref{sig.arm} we compute the
significance from 
Eq.~(\ref{eq:sig}) for
UPOs as a function of the scale size. The
shapes of the curves are 
very similar and the scales with hints for determinism are in almost all
cases the same or at least very close to each other. Furthermore we
see that this 
simple model yields scales with relatively high 
significances. While the dynamics for small scales is rather stochastic
and provides only small significances, we observe the most
deterministic behavior for intermediate scales. For large scales the
significance decreases again due to averaging effects in the dynamics,
but does not tend to zero. 

However, in contrast to the real earthquake data the model is nearly 
homogeneous in space. For more realistic models one could add a
component in Eq.~(\ref{eq:bak}), which is space--dependent.   

\begin{figure} 
\begin{center}
\begin{minipage}{0.48\textwidth}
\begin{center}
\epsfxsize=8cm
\epsfysize=6cm
\epsfbox{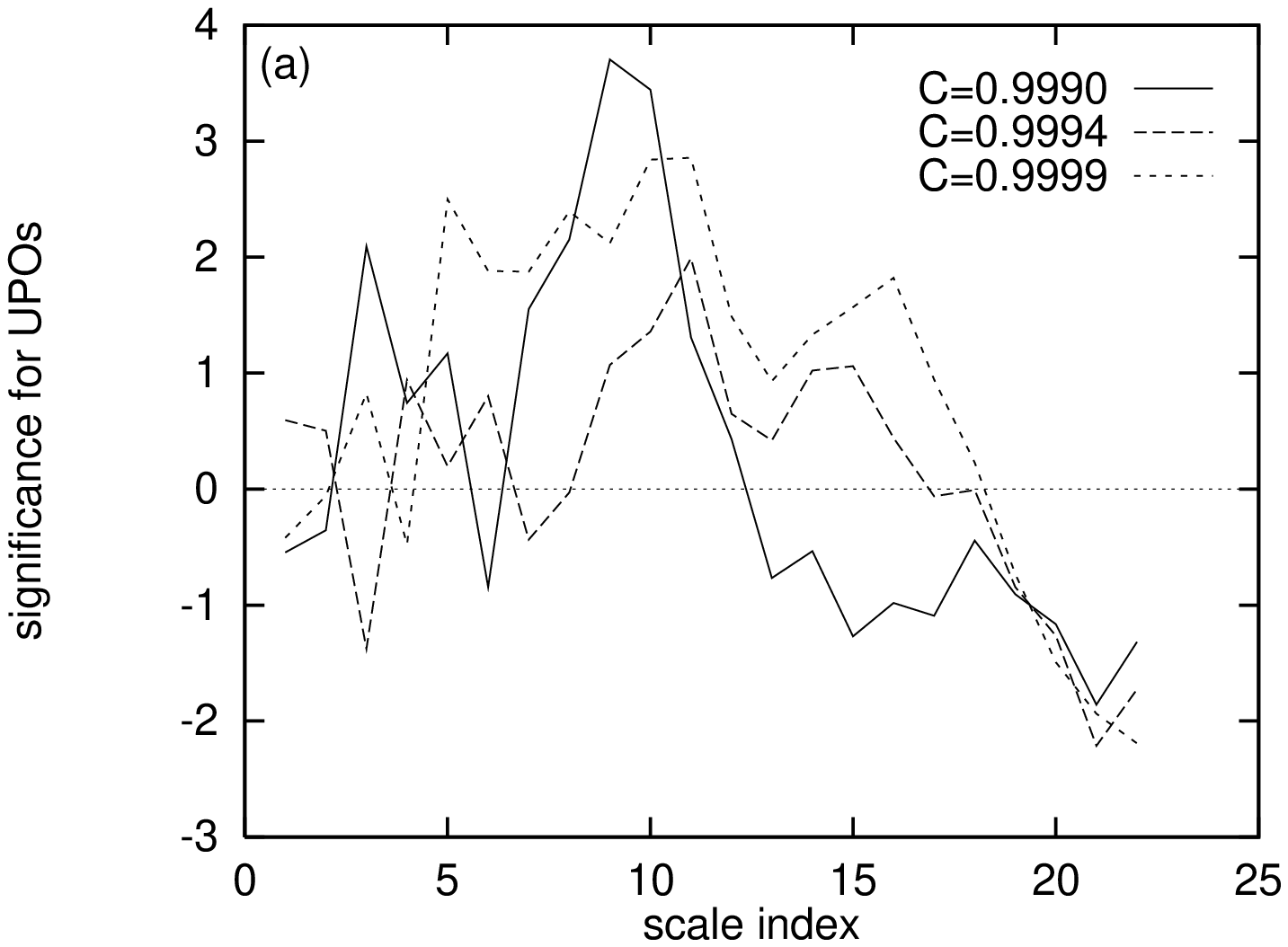}
\end{center}
\end{minipage}
\begin{minipage}{0.48\textwidth}
\begin{center}
\epsfxsize=8cm
\epsfysize=6cm
\epsfbox{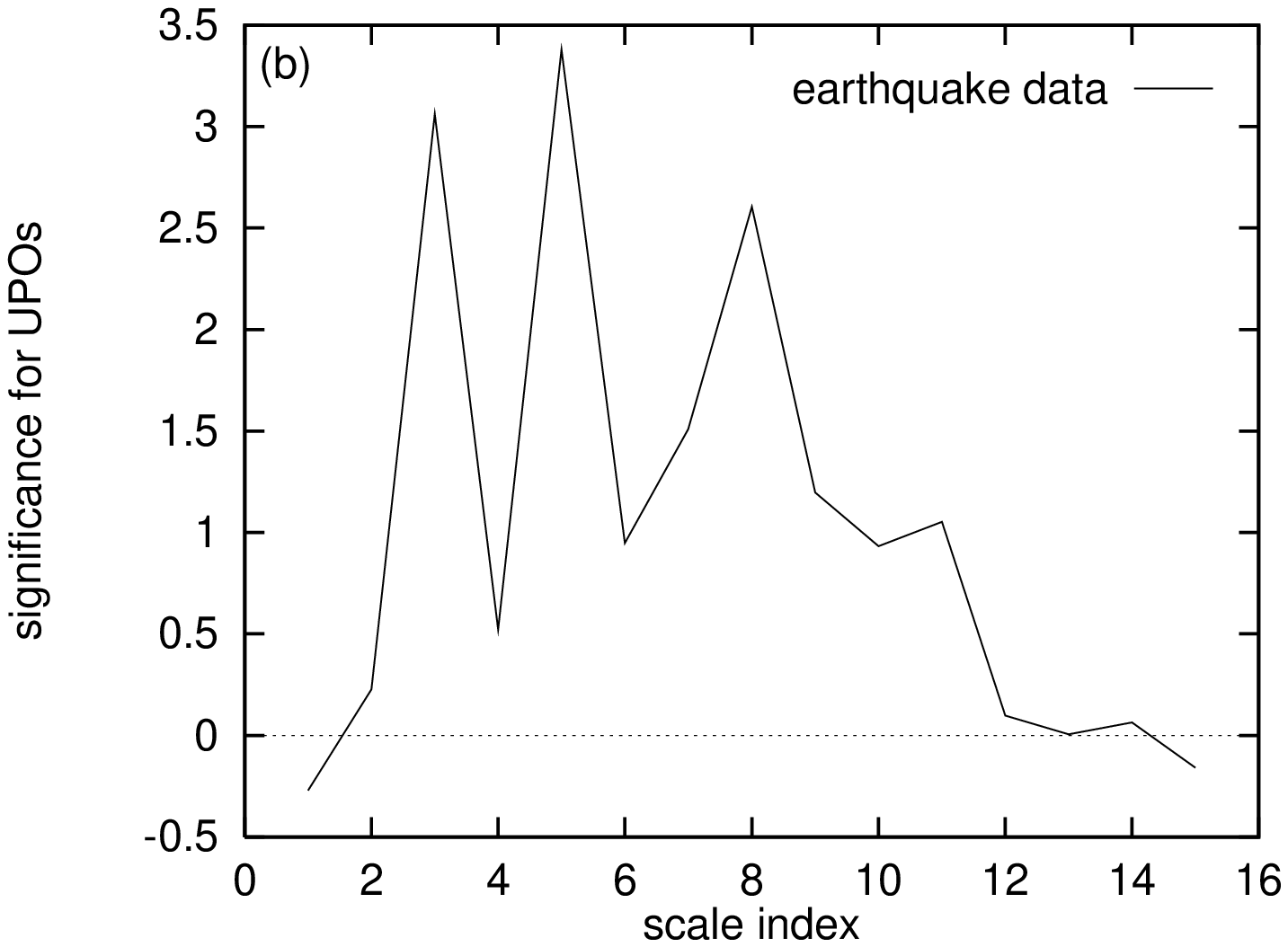}
\end{center}
\end{minipage}
\end{center}
\caption{(a) shows the significance for UPOs
depending on the scale for three different couplings 
in the model described in the text (Eq.~\protect\ref{eq:bakmod}). For each
scale the average number (over space) of UPOs is compared with 30
surrogate earthquake catalogues. The definition of the significance
is analogue to Eq.~(\ref{eq:sig}). One curve belongs to one
catalogue. (b) results from the same calculation
for the Armenia catalogue. A unit of the scale index is 10 km.}
\label{coup}
\end{figure}

In the following we study another modification of the model, which
yields a physical interpretation of the characteristic scales. 
Considering models with global coupling forces
(e.g. slider--block--models),  the size of the scale should increase
with growing coupling.  
To proof this for our model, we introduce a coupling strength $C$  
by changing the rules for the energy release. 
If $Z(i,j)>Z_{crit}$, the redistribution of the
energy follows now 
\begin{eqnarray}
\label{eq:bakmod} 
\lefteqn {Z(i,j)\longrightarrow 0}  \\
&Z(i\pm 1,j\pm 1)\longrightarrow Z(i\pm 1,j\pm 1)+(1-\epsilon)\frac{Z(i,j)}{4}
\nonumber 
\end{eqnarray}
so that an increasing value of $\epsilon$ corresponds to a decreasing
coupling strength $C:=1-\epsilon$. 
Note that $Z$ is noninteger here 
in contrast to the model in~\cite{Baketal89}. 
Moreover, a cell in a critical state decays by transferring its
total energy to the neighbor cells.
This modification is done in order to produce a large magnitude
spectrum. In the model with integer energy units, the number of
different magnitudes decreases very fast with decreasing coupling. 
For different couplings we compute the total number of UPO candidates and
compare it with 30 surrogate earthquake catalogues that are generated
by randomizing the original catalogue such 
that the distribution is conserved. For one scale the mean 
number of UPO candidates is computed. In Fig.~\ref{coup}(a) we show the
statistical significance from Eq.~(\ref{eq:sig}) for three different
couplings $C$. We observe a significant 
deviation between the model data and the mean of the surrogate data for
intermediate scales. For small and large scales no significant determinism is
present due to the statistics and averaging effects. Although the
model is driven by stochastic forces, the dynamics behaves
deterministic on certain scales. More important is the dependence of
these scales on the coupling: for strong couplings we observe still positive
significances on large scales; in contrast to this, the significance
decreases very fast in  this  scale range for small couplings. On 
Fig.~\ref{coup}(b) the results for the calculation with the
real data are given, which shows a qualitatively good agreement with
the model with strong coupling in Fig.~\ref{coup}(a).  

Due to these results from pure model studies, we get indications that
large spatial scales in real earthquake data correspond to strong
coupling forces in the earth's crust.

\section{Summary and Outlook}\label{sec:summary}
We have presented a new technique to characterize the dynamics of 
spatially extended natural systems. In particular, we have analysed earthquake
catalogues, but in principle the method can be regarded as a part
of a general approach for data analysis, because it is applicable to a
large variety of systems. 

The main idea is to check whether it exists an intermediate spatial scale
between the noisy micro-scales and the large scales, where the dynamics
is dominated by averaging effects. This intermediate scale is
then characterized by a maximum of nontrivial determinism and it
represents the appropriate length scale, on which the main 
features of the underlying dynamics can be observed.  
To extract the characteristic spatial scale from the data, we look for 
unstable periodic orbits in time series corresponding to different
scales. The occurrence of such orbits is a measure for nonlinear and
low--dimensional determinism. The scale with the highest significance 
(with respect to the condition $K\ge 3$)  -- in comparison with
surrogate data -- can be considered as the 
characteristic scale. 

Our calculations show that intermediate scales with deterministic
dynamics in the sense as mentioned above exist in earthquake data. In some
spatial regions we observe a clear maximum for the significance as a
function of the scale size. Moreover, in some regions the 
statistical significance reaches values up to seven. 

An interesting result arising from studies with model data 
is the interpretation of the spatial scales. Using a 
modification of the well--known model of Bak and Tang, we
have shown that the scale is related to 
a coupling strength, i.e. high coupling strengths correspond to large scales. 
This seems to be an interesting tool to evaluate
models, in particular such with locally varying coupling forces. 

However, one has to keep in mind that our ansatz is still very
general and should be refined for special  applications. For instance some 
effects of seismicity are not yet included 
in our approach.  Perhaps the technique can be improved
by using more complicated areas than circles. 
This would be well--adapted for the study of spatial
inhomogeneities. In this context it is a special challenge to analyse
the influence of such inhomogeneities in simple models. 

In the future we should focus on a detailed analysis of the time
series. 
From seismology we know that every main shock is more or less
accompanied by precursery phenomena~\cite{Schretal90} and
aftershocks~\cite{Moletal90}. It is an interesting question, if the occurrence
of UPOs in the time series can be connected with these patterns.

We believe, however, 
that our technique is promising for the analysis of a
large variety of spatially extended natural systems.

\acknowledgements{The authors are grateful to J.\ Zschau and his
group at the GeoForschungsZentrum Potsdam for the fruitful 
cooperation and the data.}  


%
%

%
%

\end{document}